\documentclass{iau}
\usepackage{graphicx}

\title[Zigzag Microwave Quasi-periodic Pulsations] %% give here short title %%
{An Peculiar Microwave Quasi-periodic Pulsations with Zigzag Pattern
in a CME-related Flare on 2005-01-15}

\author[Baolin Tan]   %% give here short author list %%
{Baolin Tan$^1$}

\affiliation{$^1$Key Laboratory of Solar Activity, National
Astronomical Observatories of the Chinese Academy of Sciences,
Beijing 100012, China. \\ email: {\tt bltan@nao.cas.cn} \\[\affilskip]}

\pubyear{2012}
\volume{294}  %% insert here IAU Symposium No.
\pagerange{119--126}
 \jname{Solar and Astrophysical Dynamos and
Magnetic Activity} \editors{A.G. Kosovichev, E.M. de Gouveia Dal
Pino, $\&$ Y.Yan}
\begin{document}

\maketitle

\begin{abstract}

A microwave quasi-periodic pulsation with zigzag pattern (Z-QPP) in
a solar flare on 2005-01-15 is observed by the Chinese Solar
Broadband Spectrometer in Huairou (SBRS/Huairou) at 1.10-1.34 GHz.
The zigzag pulsation occurred just in the early rising phase of the
flare with weakly right-handed circular polarization. Its period is
only several decades millisecond. Particularly, before and after the
pulsation, there are many spectral fine structures, such as zebra
patterns, fibers, and millisecond spikes. The microwave Z-QPP can
provide some kinematic information of the source region in the early
rising phase of the flare, and the source width changes from
$\sim$1000 km to 3300 km, even if we have no imaging observations.
The abundant spectral fine structures possibly reflect the dynamic
features of non-thermal particles.

%% add here a maximum of 10 keywords, to be taken form the file <Keywords.txt>
\end{abstract}

%\firstsection % if your document starts with a section, remove some space above using this command.

%\section{Introduction}

Solar microwave spectral fine structures are the most important and
interesting phenomena, which can provide many intrinsic features of
solar eruptions. This work reports first time a peculiar fine
structure: a microwave quasi-periodic pulsation (QPP) with zigzag
pattern (abbreviated as Z-QPP, hereafter) in a flare on 2005-01-15.

\begin{figure}[b]
% \vspace*{-2.0 cm}
\begin{center}
 \includegraphics[width=2.4in, height=2.4in]{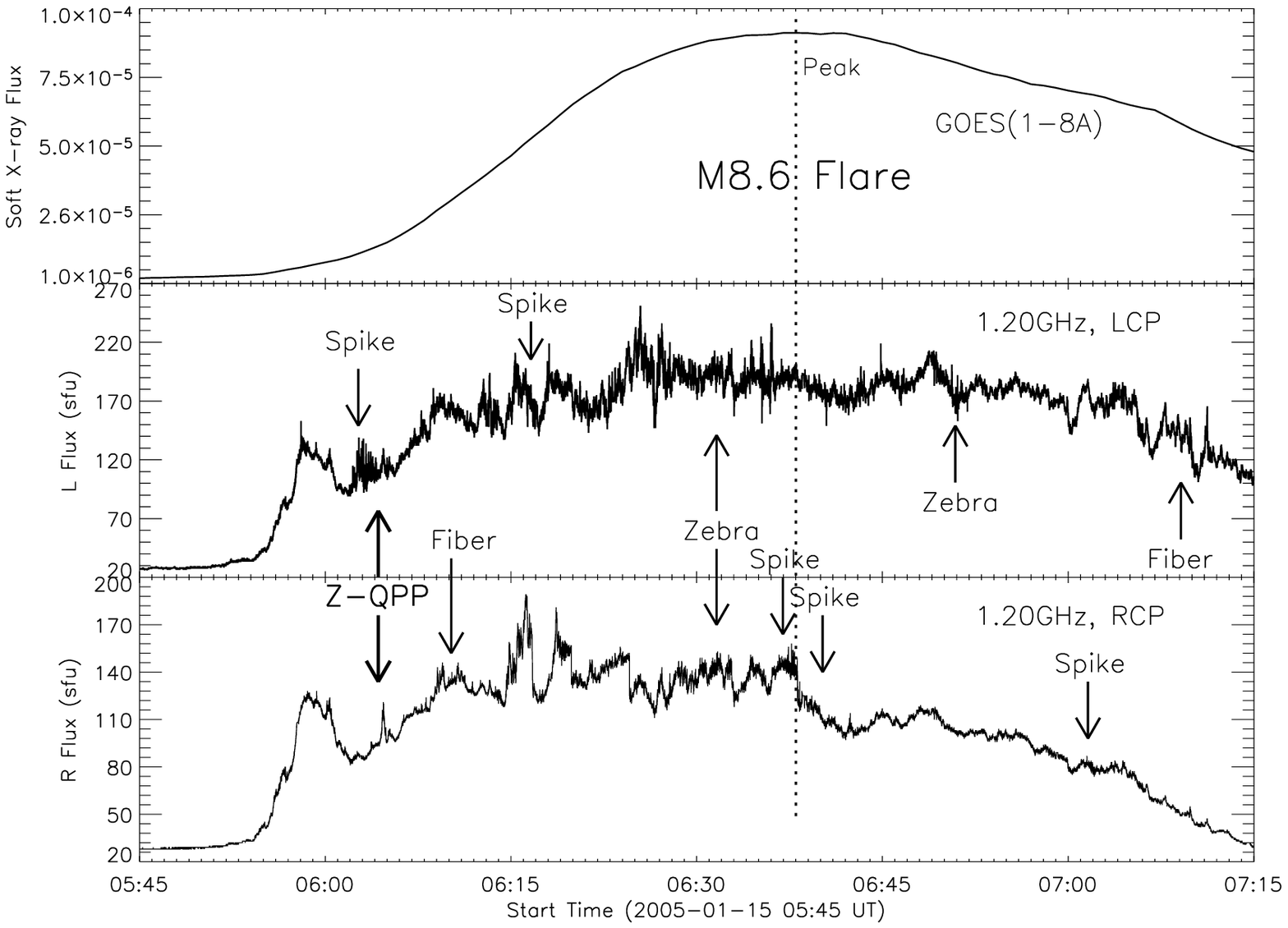}
 \includegraphics[width=2.8in, height=2.4in]{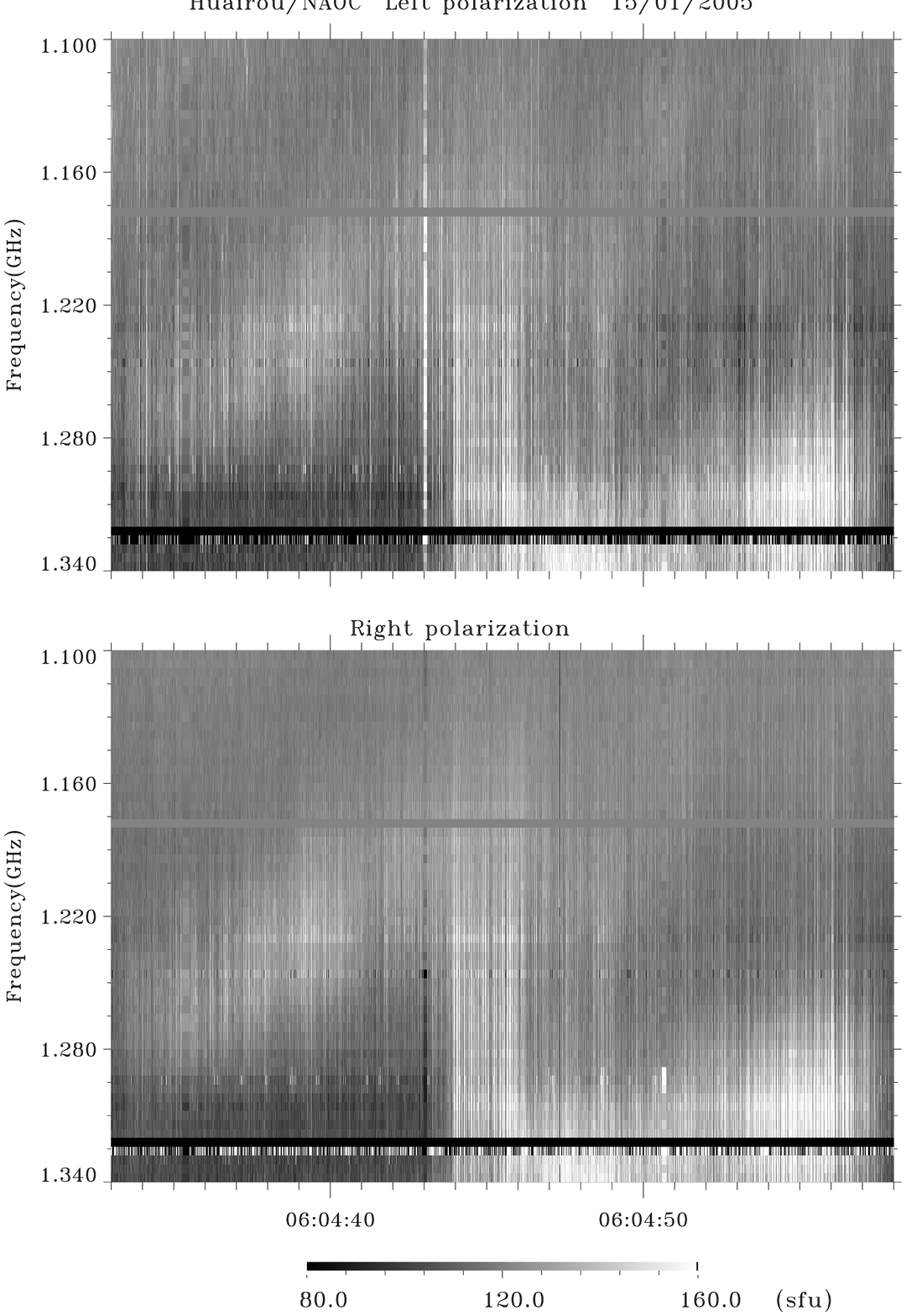}
% \vspace*{-1.0 cm}
 \caption{Left panels are the profiles of the M8.6 long-duration flare on 2005-1-15. Soft X-ray intensity observed by GOES (upper), microwave emission of
   left-handed circular polarization at 1.20 GHz (middle), and the microwave emission of right-handed circular polarization at 1.20 GHz (bottom). Right
   panels are the spectrogram of microwave quasi-periodic pulsation with zigzag pattern (Z-QPP) in the early rising phase of the flare.}
   \label{fig1}
\end{center}
\end{figure}

The flare is an M8.6 class long-duration event accompanying with a
powerful CME in active region AR10720 with location of N16E04, very
close to the center of solar disk. It is observed by the Chinese
Solar Broadband Spectrometer in Huairou (SBRS/Huairou) at frequency
of 1.10-1.34 GHz (4.0 MHz resolution, 1.25 ms cadence). The left
panels of Figure 1 show that the flare lasts from 05:54 UT to 07:17
UT, for about 83 minutes (Cheng et al, 2010). The Z-QPP occurred
just in the early rising phase of the flare. The right panels of
present the Z-QPP with left- and right-handed circular polarization
spectrogram. The Z-QPP is weakly right-handed circular polarization,
its period is only several decades of milliseconds, which belongs to
very short period pulsation. We may partition the Z-QPP into three
paragraphs. The first paragraph starts at 06:04:32 UT, ends at
06:04:43 UT, the global frequency drifting rate is about -16.5
MHz/s, the pulse frequency drifting rate is 2.30-9.60 GHz/s, the
period is 40-85 ms with averaged value of 60 ms. The bandwidth
increases slowly from 60 MHz to 160 MHz. The second paragraph starts
at 06:04:43 UT and ends at 06:04:47 UT, its global frequency
drifting rate is about 0 and the pulse frequency drifting rate is
9.60-14.40 GHz/s, the period is 48-106 ms, and the bandwidth 140-200
MHz, few variations. The third paragraph starts at 06:04:47 UT and
ends at 06:04:47 UT, its global frequency drifting rate is about
-5.6 MHz/s and the pulse frequency drifting rate is 6.7-20 GHz/s,
the period is 38-87 ms, and the bandwidth increases slowly from 60
to 140 MHz, and possibly beyond frequency domain (1.34 GHz).

The frequency distance between the two adjacent strips of the
concomitant zebra pattern is about 80 MHz, which implies that the
magnetic field in the source region is about 71.5 Gs from DPR model.
The weakly polarization and strongly emission intensity imply that
plasma emission is most possibly the emission mechanism. The Z-QPP
may reflect the variations of the physical conditions and the
dynamic processes in source region (Aschwanden \& Benz, 1986; Tan,
2008). The velocities of source motion or energetic particles ($v$),
and the width of source region (L) can be estimated:

\begin{equation}
v=2\frac{df}{fdt}H_{n},~~~~~L=2\frac{\triangle f}{f}H_{n}.
\end{equation}

$H_{n}$ is the scale length of plasma density in the background for
the source motion and in the source region for the energetic
particles.

The above estimation of the Z-QPP indicates the source region moves
in a speed from 275-93 km/s upwards, the source width expands from
1000 km to 3300 km, and the associated speeds of energetic particles
is about 0.13~0.53 $c$, 0.53-0.8 $c$, and 0.36-0.9 $c$ in the three
paragraphs, respectively. Here, $c$ is the light speed. Before and
after the Z-QPP, there are many spectral fine structures, such as
zebra patterns, fibers and millisecond spikes, etc. Some of them are
marked in Figure 1. The abundant spectral fine structures reflect
the dynamic features of the non-thermal particles (Huang \& Tan,
2012).

\acknowledgments

This work is supported by NSFC Grant No. 11273030, 10921303, MOST
Grant No. 2011CB811401, and the National Major Scientific Equipment
R\&D Project ZDYZ2009-3.

\end{document}